\documentstyle[12pt]{article}

\begin{document}

\baselineskip 6mm
\renewcommand{\thefootnote}{\fnsymbol{footnote}}

\begin{titlepage}

\hfill\parbox{4cm}
{ KIAS-P99017 \\ hep-th/9903022 \\ March 1999}

\vspace{15mm}
\begin{center}
{\LARGE \bf
Supersymmetric completion of supersymmetric
quantum mechanics}
\end{center}

\vspace{5mm}
\begin{center} 
Seungjoon Hyun\footnote{\tt hyun@kias.re.kr}, 
Youngjai Kiem\footnote{\tt ykiem@kias.re.kr} and 
Hyeonjoon Shin\footnote{\tt hshin@kias.re.kr} 
\\[5mm] 
{\it School of Physics, Korea Institute for
Advanced Study, Seoul 130-012, Korea }
\end{center}
\thispagestyle{empty}

\vfill
\begin{center}
{\bf Abstract}
\end{center}
\noindent
Via supersymmetry argument, we determine the effective action of the
$SU(2)$ supersymmetric Yang-Mills quantum mechanics up to two
constants, which results from the full supersymmetric completion of
the $F^4$ term.  The effective action, consisting of zero, two, four,
six and eight fermion terms, agrees with the known perturbative
one-loop calculations from the type II string theory and the matrix
theory.  Our derivation thus demonstrates its non-renormalization
properties, namely, the one-loop exactness of the aforementioned
action and the absence of the non-perturbative corrections.  We
briefly discuss generalizations to other branes and the comparison to
the DLCQ supergravity analysis.  In particular, our results show that
the stringent constraints from the supersymmetry are responsible for
the agreement between the matrix theory and supergravity with sixteen
supercharges.
\vspace{2cm}
\end{titlepage}

\baselineskip 7mm
\renewcommand{\thefootnote}{\arabic{footnote}}
\setcounter{footnote}{0}

\section{Introduction and Summary}

Supersymmetric matrix quantum mechanics (we alternatively call it
matrix theory or supersymmetric quantum mechanics in this paper) is
suggested to provide us with a quantum description of the
eleven-dimensional supergravity in the large $N$ limit \cite{bfss}.
For the finite $N$ case, the eleven-dimensional supergravity
formulated in terms of the discrete light-cone quantization (DLCQ)
scheme of Susskind \cite{susskind,seiberg,eleven} is argued to be
described by the supersymmetric quantum mechanics.  The agreement of
the effective action between the matrix theory and the DLCQ
supergravity for particle (or other extended objects in $M$ theory
that we do not consider in this paper) scatterings is by now
well-reported in the literature \cite{becker1}-\cite{new}.

The impressive agreement between these two radically different
theories naturally lead us to wonder why they should agree in the
first place.  Intuitively, supersymmetries should play a key role; the
scattering dynamics analyzed in, for example, Refs.~\cite{becker1} and
\cite{becker} preserves sixteen supersymmetries.  The matrix quantum
mechanics, being the dimensional reduction to one-dimension from the
ten-dimensional supersymmetric Yang-Mills (SYM) theory, describes the
low energy dynamics of the D-particles in IIA string theory
\cite{polchin} and possesses sixteen supersymmetries along with the
$SO(9)$ R-symmetry.  Similarly, the supergravity space-time metric for
$M$-momenta moving along the light-cone direction in the DLCQ
supergravity has sixteen Killing spinors and $SO(9)$ transversal
rotational isometry \cite{new}.  In the latter case, the detailed form
of the metric is determined by specifying a nine-dimensional harmonic
function, which is obtained by solving the BPS equations that are
valid when there exist sixteen unbroken supersymmetries.  Once the
metric is determined in the supergravity side, the bosonic probe
action that produces $v^4$ term in the small velocity expansion can be
straightforwardly written down \cite{becker}.  Given this purely
bosonic $v^4$ term, the supersymmetrization uniquely determines all
the other fermion terms via the superspace formalism \cite{new}.

A similar behavior to what happens in the supergravity case has been
observed in the pioneering work of Paban, Sethi and Stern
\cite{sethi2} in the matrix quantum mechanics.  They show that, in the
case of the supersymmetric quantum mechanics, the eight fermion terms,
that result from the supersymmetric completion of the one-loop $F^4$
term, can be uniquely determined (up to an overall normalization) by
the supersymmetry argument alone; the coefficient functions of the
eight fermion terms satisfy an analog of the BPS equations from the
supergravity \cite{sethi2}.  If the correspondence between the matrix
theory and the supergravity holds up as presumed, one naturally hopes
further that the eight fermion terms, once determined, should
determine all the other remaining terms with zero, two, four, and six
fermions.  In this paper, we show that the constraints from the
sixteen supersymmetries determine all fermion terms in the effective
action belonging to the full supersymmetric completion of the $F^4$
terms of the matrix quantum mechanics.  Our results demonstrate the
formal similarity of the matrix quantum mechanics to the supergravity
where the superspace formalism generates all fermion terms from the
purely bosonic terms.

Main benefit of this line of approach is that the non-renormalization
theorem \cite{sethi2,lowe} for all the terms that we calculate is
guaranteed, since the sixteen supersymmetries are the exact symmetries
in our context.  Our effective action turns out to be identical to the
one-loop perturbative terms reported in the literature including the
bosonic term \cite{becker}, two fermion terms \cite{kraus2}, four
fermion terms \cite{mcart}, and eight fermion terms \cite{barrio}.
The six fermion terms have not been calculated in the perturbative
supersymmetric quantum mechanics framework, but our results are
identical to the ones obtained by the perturbative analysis in the IIA
string theory framework \cite{harvey,morales,review}.  In view of
these, our analysis demonstrates the non-renormalization properties,
including the one-loop exactness and the lack of non-perturbative
effects, for those terms that originate from the supersymmetric
completion of the $F^4$ terms.  Specifically, following the notations
introduced in Sec.~2.1, our results are:
\begin{equation}
 I = \Gamma_{(2)} + \Gamma_{(4)} + F^6 ~ {\rm terms} 
+ \cdots ~,
\label{first}
\end{equation}
where
\begin{equation}
\Gamma_{(2)} =  \int d \lambda 
\left( \frac{1}{2} v^2  + \frac{i}{2} \theta 
 \dot{\theta} \right) ~,
\end{equation}
and
\begin{eqnarray}
\Gamma_{(4)} & = & \int d \lambda \; 
      \Big\{
	f^{(0)} (\phi ) (v^2 )^2 
   	+ v^2 v^i f^{(2)}_1 (\phi) 
 	\phi^j (\theta \gamma^{ij} \theta ) 
			\nonumber\\ 
 & &  \hspace{12mm}
	+ v^i v^j  
	\left[ 	f^{(4)}_2 (\phi ) \phi^k \phi^l
		+ f^{(4)}_0 (\phi ) \delta^{kl}  
	\right]
 	(\theta \gamma^{ik} \theta ) 
	(\theta \gamma^{jl} \theta )   
			\nonumber \\
 & &  \hspace{12mm}
	+ v^i 
	\left[ 	f^{(6)}_3
 		\phi^j \phi^k \phi^l 
		+ 2 f^{(6)}_1 \delta^{jk} \phi^l  
	\right]
 	(\theta \gamma^{ij} \theta)
  	(\theta \gamma^{km} \theta)
  	(\theta \gamma^{lm} \theta )
			\nonumber \\
 & &  \hspace{12mm}
	+ \left[ f^{(8)}_4 (\phi ) \phi^i \phi^j \phi^k \phi^l
 		+ 4 f^{(8)}_2 (\phi ) \delta^{ik} \phi^j \phi^l    
 		+ 2 f^{(8)}_0 (\phi ) \delta^{ik} \delta^{jl} 
	  \right]
 			\nonumber \\
 & &  \hspace{16mm}
	\times (\theta \gamma^{im} \theta)
	(\theta \gamma^{jm} \theta)
  	(\theta \gamma^{kn} \theta ) 
	(\theta \gamma^{ln} \theta ) 
      \Big\} ~,
\end{eqnarray}
where $f^{(0)}$ is an $SO(9)$ invariant nine-dimensional harmonic
function
\begin{equation}
f^{(0)} = k_1 + k_2 \frac{1}{\phi^7} ~,
\end{equation}
the functions $f^{(2p)}_q$ satisfy
\begin{equation}
f^{(2p)}_q = C_p \left( \frac{d}{\phi d \phi} 
 \right)^{(p+q)/2} f^{(0)} ~,
\label{ansatz}
\end{equation}
and the numbers $C_p$ independent of $q$ are given by
\begin{equation}
C_1 =  \frac{i}{2}  ~, ~ C_2 = - \frac{1}{8}  ~ , ~ 
C_3 = - \frac{i}{144} ~ , ~ C_4  = \frac{1}{8064}   ~.
\label{last}
\end{equation}
In other words, the effective action $\Gamma_{(4)}$ is fully
determined up to two constants $k_1$ and $k_2$.  These results are
obtained {\em purely} on the basis of the existence of sixteen
supersymmtries, $SO(9)$ R-symmetry and the CPT invariance.
Perturbative calculations of the bosonic effective action within the
matrix theory framework \cite{becker} show that $k_1 = 0$, while the
supersymmetry allows it to be an arbitrary constant.  To recover $k_1
= 0$ from the supergravity necessitates the use of the DLCQ framework
where the asymptotic time direction of the background geometry is
light-like \cite{eleven,becker,new}.  The resulting background
geometry is the non-asymptotically flat near-horizon D-particle
geometry in ten dimensions \cite{ten}, or equivalently, the
asymptotically flat Aichelberg-Sexl geometry in eleven dimensions
\cite{eleven}.  The asymptotic time direction of the asymptotically
flat D-particle background geometry is time-like, which implies $k_1 >
0$.  Apart from the necessity of introducing the DLCQ framework, which
is strictly speaking beyond supersymmetry argument, the complete
effective action itself is determined by the supersymmetry with
sixteen supercharges, up to an overall normalization.  Therefore, the
stringent constraints imposed by the maximal supersymmetries are
responsible for the agreement for the two-body dynamics between the
DLCQ supergravity and the matrix theory when there are sixteen
supersymmetries.  In view of this aspect, the crucial future tests to
verify the (dis)agreements between the matrix theory and the
supergravity should be directed to the cases when some of the
supersymmetries are broken, as well as to the cases involving
multi-body (especially large $N$) scatterings \cite{newsusskind}.

The technical details are presented in Sec.~2, and we briefly discuss
related issues, such as the extension of our analysis to the membrane
case, in Sec.~3.

\section{Supersymmetric completion of $F^4$ terms}

This section is organized as follows.  In Sec.~2.1, we classify the
terms that can appear in the effective action obtained by the
supersymmetric completion of the $F^4$ terms.  The guiding principles
here are the unbroken $SO(9)$ R-symmetry and the CPT invariance, both
of which constrain the possible terms in the effective action.
Appendix A and B contain some technical details such as a number of
relevant Fierz identities.  In Sec.~2.2, starting from the effective
action constrained in Sec.~2.1, we explicitly work out the
supersymmetry transformations to determine the full supersymmetrically
completed effective action.

\subsection{Constraints from the $SO(9)$ $R$-symmetry
and CPT on the effective action}

The effective action of the SYM quantum mechanics is described by nine
scalars $\phi^i$ and their time derivatives $v^i = (d / d \lambda )
({\phi}_i ) \equiv \dot{\phi}^i$ where $i = 1 , \cdots , 9$ and
$\lambda$ is a time coordinate.  At the origin of the SYM moduli
space, the $R$-symmetry is unbroken $SO(9)$.  This is the situation
that corresponds to the matrix theory description of the source-probe
two-body dynamics where the probe $M$-momentum moves in the background
geometry of the $N$ coincident source $M$-momenta.  In this paper, we
will start by computing the constraints imposed on the possible terms
in the effective action resulting from the requirement of the unbroken
$SO(9)$ $R$-symmetry.  Furthermore, we will impose the CPT invariance
that the matrix quantum mechanics inherits from the ten-dimensional
IIA SYM theory.  Our starting point will be the consideration of the
$F^2$ terms, following the notation and set-up of Ref.~\cite{sethi2}.
Considering the $SO(9)$ invariance, the possible bosonic $F^2$ terms
are $g_1 (\phi ) v^2 $ and $g_2 (\phi ) (v^i \phi^i )^2 $.  Here the
$SO(9)$ vector indices are contracted by the Kronecker delta
$\delta_{ij}$, for example, $v^2 = \delta_{ij} \dot{\phi}^i
\dot{\phi}^j$ , and $g_1$ and $g_2$ are $SO(9)$ invariant scalar
functions, which depend only on an $SO(9)$ invariant $\phi =
\sqrt{\phi^i \phi^i }$.  By the inverse of the diffeomorphism of the
form
\begin{equation}
h_1 (\phi ) v^i \rightarrow  h_2 (\phi )  v^i +  h_3 
  (\phi ) (v^j \phi^j ) \phi^i ~,
\label{diffeo}
\end{equation}
which squares to
\begin{equation}
h_1^2 v^2 \rightarrow  h_2^2 v^2 + 
( 2 h_2  h_3  + h_3^2
 \phi^2  )  ( v^j \phi^j )^2 \equiv
   g_1   v^2 + g_2  (v^i \phi^i )^2 ~,
\label{dsq}
\end{equation}
the general $SO(9)$ invariant moduli space metric consisting of the
two terms $g_1 v^2 d \lambda^2$ and $g_2 ( v^i \phi^i )^2 d \lambda^2
$ can be reduced to a simple form as follows
\begin{equation}
 ds^2 = g_1  d \phi^i d \phi^i + g_2 \phi^i \phi^j
  d \phi^i d \phi^j = h_1^2 d \phi^i d \phi^i ~.
\end{equation}
The key result of Ref.~\cite{sethi2} is that this moduli space metric
is constrained to be flat (corresponding to a free abelian theory)
under the imposition of the supersymmetry with sixteen supercharges.
For the $F^2$ terms, we thus choose a coordinate such that the
quadratic effective action looks like the following form
\begin{equation}
\label{v2act}
\Gamma_{(2)} = \int d \lambda \;
\left( 	\frac{1}{2} v^2  + \frac{i}{2} \theta 
  	\dot{\theta} 
\right) ~,
\end{equation}
which is invariant under the supersymmetry transformation
\begin{equation}
 \delta \phi^i  = - i \epsilon \gamma^i \theta 
\label{quadsusy}
\end{equation}
\[ \delta \theta = (\gamma^i v^i \epsilon ) . \]
Here $\theta$ is the sixteen component $SO(9)$ Majorana spinor and
gamma matrices $\gamma^i$ are the $16 \times 16$ SO(9) gamma matrices.
For the following description, we introduce an ordering $O( v^i ) = O (
\theta \theta ) = O( d / d\lambda ) =1 $ and we assign $O(\epsilon ) =
-1/2$ for the $SO(9)$ Majorana spinor parameter $\epsilon$ for the
supersymmetry transformation.  The subscript of the effective action
$\Gamma_{(2)}$ signifies the fact that the terms of (\ref{v2act}) are
of order two under the above ordering assignment.

What we are interested in in this paper is to determine the order of
four terms, $\Gamma_{(4)}$, in the effective action when the spinor
$\theta$ and the velocity $v^i$ are constant\footnote{These are
consistent with the supersymmetry transformation,
Eq.~(\ref{quadsusy}).  As will be explained in Sec.~2.2, the inclusion
of $\Gamma_{(4)}$ modifies the supersymmetry transformation,
Eq.~(\ref{quadsusy}).  The detailed consideration in Sec.~2.2 will
show that using the leading order supersymmetry transformation,
Eq.~(\ref{quadsusy}), is enough for $\Gamma_{(4)}$, while the
correction plays an important role for the supersymmetric variation of
$\Gamma_{(2)}$.}.  In this case, we can preclude the possible
acceleration and high order fermion derivative terms, which are, in
general, present in the effective action.  Due to the existence of the
sixteen supersymmetries and the ordering assignment in the above, the
structure of $\Gamma_{(4)}$ can be schematically written as
\cite{harvey,morales}
\begin{equation}
\Gamma_{(4)} = \int d \lambda 
\left( \left[ v^4 \right] + \left[ v^3 \theta^2 \right]
+ \left[ v^2 \theta^4 \right] + \left[ v \theta^6 \right]  
+ \left[ \theta^8 \right] \right) ~,
\label{gamma4}
\end{equation}
where we suppress index structures; it consists of zero, two, four,
six and eight fermion terms.  The possible terms that can appear in
Eq.~(\ref{gamma4}) can be constrained by the requirement of the
unbroken $SO(9)$ $R$-symmetry and the CPT theorem, as we will discuss
now.  Since $\theta$ is an $SO(9)$ Majorana spinor, the fermion
bilinears satisfy $\theta \gamma^{i_1 i_2 \cdots i_k} \theta = 0$ for
$k = 0, 1, 4, 5, 8, 9$.  Here $\gamma^{i_1 i_2 \cdots i_k}$ is the
totally anti-symmetrized $k$-product of gamma matrices normalized to
unity.  Therefore, the fermion structure of the $p$-fermion term shown
in Eq.~(\ref{gamma4}) is in general a $(p/2)$-product of fermion
bilinears $J^{ij} \equiv \theta \gamma^{ij} \theta$ and $K^{ijk}
\equiv \theta \gamma^{ijk} \theta$.  The terms appearing in
Eq.~(\ref{gamma4}) can thus be classified as shown in Table 1.  
\\
\begin{center}
\begin{tabular}{|c|c|c|c|c|c|} \hline
      0-fermion  &  $4^{v^4}$ & & & & \\ \hline
      2-fermion  & $5^{v^3 J}$  & $6^{v^3 K}$ & & & \\ \hline
      4-fermion  & $6^{v^2 JJ}$ & $7^{v^2 JK}$
                     &  $8^{v^2 KK}$ & & \\ \hline
     6-fermion  & $7^{v JJJ}$   & $8^{vJJK}$ &
                $9^{vJKK}$  & $10^{vKKK}$ & \\ \hline
     8-fermion  &  $8^{JJJJ}$ & $9^{JJJK}$ & $10^{JJKK}$ 
    & $11^{JKKK}$ & $12^{KKKK}$  \\ \hline
\end{tabular} 
\\
[.3cm]
{\small Table 1. The classification of the possible terms in the
effective action. }
\end{center}
The superscript $v^m J^n K^p$ denotes the fact that the corresponding
terms are composed of the products of $m$ velocity vector $v^i$'s, $n$
fermion bilinear $J^{ij}$'s and $p$ fermion bilinears $K^{ijk}$'s.
The number $r$ ($r = 4 , \cdots , 12$) in each entry denotes the total
number of indices given by $r = m + 2 n + 3 p$.  Since the terms in
the effective action should be an $SO(9)$ scalar, an object with
indices should be contracted with an appropriate number of $\phi^i$'s;
an object with $r$ indices can however be self-contracted $s$ times ($
0 \le s \le \left[ r/2 \right]$) before the contraction with $(r-2s)$
$\phi^i$'s.  Each term in Table 1 contains, in addition to the $v^m
J^n K^p \phi^{r - 2s}$ structure, an arbitrary coefficient function as
an overall factor that depends only on an $SO(9)$ invariant $\phi =
\sqrt{ \phi^i \phi^i}$.

The CPT invariance dictates that the terms of the second ($6^{v^3 K},
7^{v^2 JK}, 8^{vJJK}, 9^{JJJK}$) and the fourth ($10^{vKKK},
11^{JKKK}$) columns of Table 1 should vanish; the terms in the first
column, which are of the form of the perturbative terms reported in
the literature, are all CPT invariant, and replacing one $J^{ij}$ with
$K^{mnp}$ turns the aforementioned terms CPT violating.  As proven in
Ref.~\cite{sethi2} for the fifth row and in Appendix A for the other
rows, all the terms in the third and fifth columns can be
Fierz-rearranged into the terms of the first column.  Therefore, we
can concentrate on the terms of the first column from now on, without
losing generality.  We note that the CPT violating terms, the terms of
the second and the fourth columns, can {\em not} be turned into the
terms of the first column via the Fierz rearrangement.  Noting a Fierz
identity $(\theta \gamma^{ij} \theta )(\theta \gamma^{ij} \theta )=
0$, all the non-vanishing terms of Table 1 can be written down as
follows, after working out all possible contractions:
\begin{eqnarray}
{\rm 0-fermion ~ :} &  (v^2 )^2 ~, \label{zerof} \\
{\rm 2-fermion ~ :} &   v^2 v^i \phi^j (\theta \gamma^{ij} 
             \theta ) ~,
 \label{twof} \\
 {\rm 4-fermion ~ :} & v^2 \phi^i \phi^j ( \theta 
     \gamma^{ik} \theta ) (\theta \gamma^{jk} 
     \theta ) ~ , \label{fourf1} \\
 &    v^i v^j \phi^k \phi^{l} (\theta \gamma^{ik} 
     \theta ) ( \theta \gamma^{jl} \theta )  ~ , 
\label{fourf2} \\
& v^i v^j (\theta \gamma^{ik} \theta ) (\theta 
    \gamma^{jk} \theta ) ~ ,                      
\label{fourf3} \\
 {\rm 6-fermion ~ :} & v^i \phi^j \phi^k \phi^l 
  (\theta \gamma^{ij} \theta)
  (\theta \gamma^{km} \theta)
  (\theta \gamma^{lm} \theta ) ~ , \label{sixf1} \\
 & v^i \phi^l 
 (\theta \gamma^{ij} \theta)
  (\theta \gamma^{jk} \theta)
  (\theta \gamma^{kl} \theta ) ~ , \label{sixf2} \\
 {\rm 8-fermion ~ :} & \phi^i \phi^j \phi^k \phi^l
  (\theta \gamma^{im} \theta)
  (\theta \gamma^{jm} \theta)
  (\theta \gamma^{kn} \theta ) 
  (\theta \gamma^{ln} \theta ) ~ , \label{eightf1} \\
&  \phi^i \phi^m (\theta \gamma^{ij} \theta)
  (\theta \gamma^{jk} \theta)
  (\theta \gamma^{kl} \theta ) 
  (\theta \gamma^{lm} \theta ) ~ , \label{eightf2} \\
& (\theta \gamma^{ij} \theta)
  (\theta \gamma^{jk} \theta)
  (\theta \gamma^{kl} \theta ) 
  (\theta \gamma^{li} \theta ) ~ , \label{eightf3}
\end{eqnarray}
and
\begin{eqnarray}
{\rm 0-fermion ~ :} &  (v^i \phi^i)^2 v^2 ~ , ~
         ( v^i \phi^i )^4  ~ , \label{szerof} \\
{\rm 2-fermion ~ :} &  (v^i \phi^i )^2  v^j \phi^k 
(\theta \gamma^{jk}  \theta ) ~ ,
 \label{stwof} \\
 {\rm 4-fermion ~ :} &  (  v^i \phi^i ) v^j \phi^{k} 
    (\theta \gamma^{jl} 
     \theta ) ( \theta \gamma^{kl} \theta ) ~ ,  
\label{sfourf1} \\
    & ( v^i \phi^i )^2  \phi^j \phi^k 
  ( \theta  \gamma^{jl} \theta ) (\theta \gamma^{kl} 
     \theta ) ~ , \label{sfourf2} \\
  {\rm 6-fermion ~ :} & (v^i \phi^i ) \phi^j \phi^m  
  (\theta \gamma^{jk} \theta)
  (\theta \gamma^{kl} \theta)
  (\theta \gamma^{lm} \theta ) \label{ssixf} ~ .
 \end{eqnarray}

According to the perturbative calculations for zero, two, four and
eight fermion terms and the type II side calculations
\cite{taylor,kraus2}\cite{mcart}-\cite{review}, the terms of the form
(\ref{szerof})-(\ref{ssixf}) do not appear in the effective action
under the choice of $\Gamma_{(2)}$ in Eq.~(\ref{v2act}).  Considering
the non-renormalization theorem of Refs.~\cite{sethi2} and
\cite{lowe}, we can set the coefficient functions of
(\ref{szerof})-(\ref{ssixf}) as zero.  In fact, consistent with the
analysis of Ref.~\cite{sethi2}, the diffeomorphism of the form
Eq.~(\ref{diffeo}) generates all the terms of
(\ref{szerof})-(\ref{ssixf}) from (\ref{zerof})-(\ref{sixf2}), just
like the same diffeomorphism generates the $(v^i \phi^i )^2$ term from
the bosonic kinetic term $v^2$ in Eq.~(\ref{v2act}) (see
Eq.~(\ref{dsq})).  Specifically, under the diffeomorphism
Eq.~(\ref{diffeo}), the terms of (\ref{zerof})-(\ref{sixf2}) generate
the following terms:
\[ (\ref{zerof}) \rightarrow  (\ref{zerof}) + 
       (\ref{szerof}) ~ , \]
\[  (\ref{twof}) \rightarrow  (\ref{twof}) +
    (\ref{stwof}) ~ , \]
\[ (\ref{fourf1}) \rightarrow (\ref{fourf1}) +
      (\ref{sfourf2}) ~ , ~ 
 (\ref{fourf2}) \rightarrow (\ref{fourf2}) ~ , ~ 
 (\ref{fourf3}) \rightarrow (\ref{fourf3})
    + (\ref{sfourf1}) + (\ref{sfourf2}) ~ , \]
\[ (\ref{sixf1}) \rightarrow (\ref{sixf1} ) ~ , ~
 (\ref{sixf2}) \rightarrow (\ref{sixf2} ) + (\ref{ssixf}) ~ , \] 
where we use the identity $\phi^i \phi^j \theta \gamma^{ij} \theta =
0$.  It is instructive to observe the same situation in the
eleven-dimensional DLCQ supergravity.  In Ref.~\cite{new}, it is shown
that (\ref{v2act}), (\ref{zerof}) and (\ref{twof}) terms are correctly
reproduced from the probe action of a massless eleven-dimensional
superparticle moving in the background geometry produced by $N$ source
$M$-momenta\footnote{The calculations for the higher fermion terms
from the supergravity side are not yet available in the literature,
except for the four fermion terms of Ref.~\cite{taylor}.}.  In the
same reference, choosing a static gauge $( d X^0 / d \lambda ) = 1$
renders the kinetic terms be of the form of (\ref{v2act}) and the
order of four terms be of the form (\ref{zerof}) + (\ref{twof}), while
the terms of the form $(v^i \phi^i)^2$, (\ref{szerof}) and
(\ref{stwof}), are absent.

We are now left to consider the terms of
(\ref{zerof})-(\ref{eightf3}).  We note the following property for the
terms of (\ref{fourf2})-(\ref{eightf3}); replacing $\phi^i \phi^j$
with $\delta^{ij}$ reduces (\ref{fourf2}) into (\ref{fourf3}),
(\ref{sixf1}) into (\ref{sixf2}), (\ref{eightf1}) into
(\ref{eightf2}), and (\ref{eightf2}) into (\ref{eightf3}), again
noting a Fierz identity $(\theta \gamma^{ij} \theta ) (\theta
\gamma^{ij} \theta )=0$.  The same replacement, when applied to
(\ref{fourf1}), makes it vanish.  Utilizing this property, the terms
of $\Gamma_{(4)}$ can in general be written as
\begin{eqnarray}
\left[ v^4 \right] &=& f^{(0)} (\phi ) (v^2 )^2 ~ ,
                         \label{f0} \\
\left[ v^3 \theta^2 \right] &=& v^2 v^i f^{(2)}_1 (\phi) 
 \phi^j (\theta \gamma^{ij} \theta ) ~ ,
                         \label{f2} \\
 \left[ v^2 \theta^4 \right] &=& v^2 g^{(4)}_2 (\phi )
  \phi^i \phi^j (\theta \gamma^{ik} \theta )
 (\theta \gamma^{jk} \theta )
			\nonumber \\
           & &  + v^i v^j  
		\left[ f^{(4)}_2 (\phi ) \phi^k \phi^l
		 + f^{(4)}_0 (\phi ) \delta^{kl}  \right]
 		(\theta \gamma^{ik} \theta ) 
		(\theta \gamma^{jl} \theta )  ~ , 
			\label{f4} \\
\left[ v \theta^6 \right] &=& v^i \left[ f^{(6)}_3
 	\phi^j \phi^k \phi^l + 2 f^{(6)}_1 \delta^{jk}
 	\phi^l  \right]
 	(\theta \gamma^{ij} \theta)
  	(\theta \gamma^{km} \theta)
  	(\theta \gamma^{lm} \theta ) ~ ,
			\label{f6} \\
\left[ \theta^8 \right] &=& 
	\left[ f^{(8)}_4 (\phi ) \phi^i \phi^j \phi^k \phi^l
 	+ 4 f^{(8)}_2 (\phi ) \delta^{ik} \phi^j \phi^l    
 	+ 2 f^{(8)}_0 (\phi ) \delta^{ik} \delta^{jl} \right]
			\nonumber \\
      & &  \times (\theta \gamma^{im} \theta)
		(\theta \gamma^{jm} \theta)
  		(\theta \gamma^{kn} \theta ) 
		(\theta \gamma^{ln} \theta ) ~ . 
			\label{f8}
\end{eqnarray}
The scalar function $f^{(2p)}_q$ represents the coefficient function
of the $q$-scalar term\footnote{Throughout this paper, the scalar
number refers to the number of scalars $\phi^i$ contracted to the
indices of the fermion bilinears.  Thus, $v^i \phi^i $, for example,
has the scalar number zero.} among the $2p$-fermion terms.  Among
$2p$-fermion terms, the maximum scalar number is $p$, as can be seen
from (\ref{f0})-(\ref{f8}).  In Sec.~2.2, we will determine the ten
coefficient functions $f^{(0)}$, $f^{(2)}_1$, $f^{(4)}_2$,
$f^{(4)}_0$, $f^{(6)}_3$, $f^{(6)}_1$, $f^{(8)}_4$, $f^{(8)}_2$,
$f^{(8)}_0$, and $g^{(4)}_2$ by the supersymmetry argument, and it
turns out that $g^{(4)}_2 = 0$.

We make the following formal observation; for a function $f$ depending
only on an $SO(9)$ invariant $\phi$, the derivatives respect to
$\phi_i$ can be computed as follows via the chain rule:
\begin{eqnarray*}
\partial_i f &=& \phi^i \left(\frac{d}{\phi d \phi}
		 \right) f ~ , \\
\partial_i \partial_j f &=&
	\phi^i \phi^j \left(\frac{d}{\phi d \phi}
 	\right)^2  f + \delta^{ij} 
 	\left(\frac{d}{\phi d \phi} \right) f  ~ , \\
\partial_i \partial_j \partial_k  f &=& 
	\phi^i \phi^j \phi^k 
	\left(\frac{d}{\phi d \phi} \right)^3 f +
	(\delta^{ij} \phi^k + \delta^{ik} \phi^j
	+ \delta^{jk} \phi^i ) \left(\frac{d}{\phi d \phi}
	\right)^2  f ~ , \\
\partial_i \partial_j \partial_k \partial_l f &=& 
	\phi^i \phi^j \phi^k \phi^l 
	\left(\frac{d}{\phi d \phi} \right)^4 f \\
      & & + ( \delta^{ij}\phi^k \phi^l +
		\delta^{ik} \phi^j \phi^l
		+ \delta^{jk}\phi^i \phi^l +\delta^{kl} 
		  \phi^i \phi^j
		+ \delta^{jl}\phi^i \phi^k +\delta^{il} 
		  \phi^j \phi^k )
		\left(\frac{d}{\phi d \phi} \right)^3 f \\
      & & + ( \delta^{ij} \delta^{kl} +
	   \delta^{ik} \delta^{jl} + \delta^{il} \delta^{jk} )
	   \left(\frac{d}{\phi d \phi} \right)^2  f ~ .
\end{eqnarray*}
Therefore, we have
\begin{eqnarray}
\partial_j f (\theta \gamma^{ij} \theta ) &=&
	\phi^j \left(\frac{d}{\phi d \phi}
	\right) f (\theta \gamma^{ij} \theta ) ~ ,
\label{ff1} \\
\partial_k \partial_l f
(\theta \gamma^{ik} \theta ) (\theta \gamma^{jl} \theta ) &=&
	\left[ \phi^k \phi^l
	\left(\frac{d}{\phi d \phi} \right)^2 f 
	+ \delta^{kl} \left(\frac{d}{\phi d \phi} \right) f  
	\right] (\theta \gamma^{ik} \theta ) 
	(\theta \gamma^{jl} \theta ) ~ ,
\label{ff2} 
\end{eqnarray}
\begin{eqnarray}
\lefteqn{ \partial_j \partial_k \partial_l f (\theta \gamma^{ij} \theta)
(\theta \gamma^{km} \theta) (\theta \gamma^{lm} \theta ) =} \hspace{20mm}
			\nonumber \\
 & & \left[ \phi^j \phi^k \phi^l 
 	\left(\frac{d}{\phi d \phi} \right)^3 f
	+ 2  \delta^{jk}
  	\phi^l \left(\frac{d}{\phi d \phi} \right)^2 f  \right]
  (\theta \gamma^{ij} \theta)
  (\theta \gamma^{km} \theta)
  (\theta \gamma^{lm} \theta ) ~ , 
\label{ff3} \\
\lefteqn{ \partial_i \partial_j \partial_k \partial_l f 
(\theta \gamma^{im} \theta)
  (\theta \gamma^{jm} \theta)
  (\theta \gamma^{kn} \theta ) 
  (\theta \gamma^{ln} \theta ) = } \hspace{23mm} 
			\nonumber \\
& & \left[  \phi^i \phi^j \phi^k \phi^l
\left(\frac{d}{\phi d \phi} \right)^4 f 
 + 4 \delta^{ik} \phi^j \phi^l \left(\frac{d}
 {\phi d \phi} \right)^3 f 
 + 2 \delta^{ik} \delta^{jl} \left(\frac{d}
{\phi d \phi} \right)^2 f \right]  \nonumber \\
& &  \times (\theta \gamma^{im} \theta)
  (\theta \gamma^{jm} \theta) 
  (\theta \gamma^{kn} \theta ) (\theta
  \gamma^{ln} \theta ) ~ , 
\label{ff4}
\end{eqnarray}
where we note that Eqs.~(\ref{ff1})-(\ref{ff4}) are identical to
Eqs.~(\ref{f2})-(\ref{f8}) if we choose
\[ f^{(2p)}_q = C_p  
 \left(\frac{d}
{\phi d \phi} \right)^{( p + q) /2  } f ~ , \]
$g^{(4)}_2 = 0$, and $C_p$ are constants.

\subsection{Supersymmery transformation and the
determination of the coefficient functions}

   From Sec.~2.1, we have explicit form of the possible terms of
$\Gamma_{(4)}$.  Upon adding $\Gamma_{(4)}$ to the quadratic terms of
$\Gamma_{(2)}$, the supersymmetry transformation law in
Eq.~(\ref{quadsusy}) should be modified.  We denote the
$\Gamma_{(4)}$-corrected supersymmetry transformation as
\begin{equation}
 \delta \phi^i =  -i \epsilon \gamma^i \theta  
 + \epsilon N^i \theta ,
\label{susy}
\end{equation}
\[ \delta \theta  =  \gamma^i v^i \epsilon 
+ M \epsilon ~. \] 
We note that $O(N)=2$ and $O(M)=3$, which let us
schematically write
\begin{equation}
N^i = \left[ v^2 N^{i(0)} \right] + 
 \left[ v N^{i(2)}  \theta^2 \right]
   + \left[ N^{i(4)} \theta^4 \right]
\label{Nexp}
\end{equation}
and
\begin{equation}
M = \left[ v^3 M^{(0)} \right] + 
 \left[ v^2 M^{(2)} \theta^2 \right]
+ \left[ v M^{(4)}  \theta^4 \right]
 + \left[ M^{(6)}  \theta^6 \right] ~.
\label{Mexp}
\end{equation}
When we take the supersymmetry variation of $\Gamma_{(2)} +
\Gamma_{(4)}$, the supersymmetry transformation Eq.~(\ref{quadsusy})
leaves $\Gamma_{(2)}$ terms invariant.  However, the correction terms
in Eq.~(\ref{susy}) generate fourth order terms from $\Gamma_{(2)}$.
Up to an order of four terms, when it comes to $\Gamma_{(4)}$ part,
considering the variation of $\Gamma_{(4)}$ under Eq.~(\ref{quadsusy})
is enough.  The correction terms in Eq.~(\ref{susy}) when acting on
$\Gamma_{(4)}$ produce terms of order six.

The variation $\delta ( \Gamma_{(2)} + \Gamma_{(4)} )$ contains one,
three, five, seven and nine $\theta$ terms, and they have to
separately vanish (up to total derivatives) for the invariance of the
effective action under supersymmetry transformations.  Specifically,
we have:
\begin{equation}
\delta_F ( \left[ v^3 \theta^2 \right] )
+ \delta_B ( \left[ v^4 \right] ) 
+ v^i  \epsilon \left[ v^2 \dot{N}^{i(0)} 
  \right] \theta
+ \frac{i}{2}  \theta \left[ v^3 \dot{M}^{(0)} \right]
  \epsilon
 \simeq 0 ~,
\label{tonef}
\end{equation}
\begin{equation}
\delta_F ( \left[ v^2 \theta^4 \right]  )
+ \delta_B ( \left[ v^3  \theta^2 \right]   ) 
+ v^i \epsilon \left[ v \dot{N}^{i(2)} \theta^2 \right] 
  \theta
+ \frac{i}{2} \theta \left[ v^2 \dot{M}^{(2)} 
  \theta^2 \right] \epsilon
 \simeq 0 ~,
\label{tthreef}
\end{equation}
\begin{equation}
 \delta_F ( \left[ v \theta^6 \right] )
+ \delta_B (  \left[ v^2 \theta^4 \right]  ) 
+ v^i  \epsilon \left[ \dot{N}^{i(4)} \theta^4 
   \right] \theta 
+ \frac{i}{2}  \theta \left[ v \dot{M}^{(4)} 
  \theta^4 \right] \epsilon
 \simeq 0 ~,
\label{tfivef}
\end{equation}
\begin{equation}
 \delta_F ( \left[ \theta^8 \right]  )
+ \delta_B ( \left[ v \theta^6 \right] ) 
 + \frac{i}{2}  \theta \left[  \dot{M}^{(6)} 
  \theta^6 \right] \epsilon
 \simeq 0 ~,
\label{tsevenf}
\end{equation}
\begin{equation}
\delta_B (  \left[ \theta^8 \right]    )  
 \simeq 0 ~, 
\label{ninef}
\end{equation}
where $\delta_B$ and $\delta_F$ represent the supersymmetric variation
of the bosonic fields and the fermionic fields, respectively.  The
symbol $\simeq$ denotes the fact that the equality holds up to a total
derivative.  We can rewrite Eqs.~(\ref{tonef})-(\ref{tsevenf}) for
an easier tractability by introducing $16 \times 16$ matrices
\begin{equation}
 L^{(p)}  = - \frac{i}{2} 
  \left[ v^{3- p/2} M^{T(p)} \theta^2 \right]
  + v^i \left[ v^{2 - p/2 } N^{i (p) } \right] ~ , 
\label{defl}
\end{equation}
where $p = 0, 2, 4, 6$.  In terms of $L^{(p)}$, 
Eqs.~(\ref{tonef})-(\ref{tsevenf}) become:
\begin{equation}
\delta_F ( \left[ v^3 \theta^2 \right] )
+ \delta_B ( \left[ v^4 \right] ) 
+  v^i \frac{\partial}
{\partial \phi^i } \left( \epsilon L^{(0)} \theta
 \right) = 0 ~ ,
\label{onef}
\end{equation}
\begin{equation}
\delta_F ( \left[ v^2 \theta^4 \right]  )
+ \delta_B ( \left[ v^3  \theta^2 \right]   )
+  v^i \frac{\partial}
{\partial \phi^i } \left( \epsilon L^{(2)} \theta
 \right) = 0 ~ ,
\label{threef}
\end{equation}
\begin{equation}
 \delta_F ( \left[ v \theta^6 \right] )
+ \delta_B (  \left[ v^2 \theta^4 \right]  )
+  v^i \frac{\partial}
{\partial \phi^i } \left( \epsilon L^{(4)} \theta
 \right) = 0 ~ ,
\label{fivef}
\end{equation}
\begin{equation}
 \delta_F ( \left[ \theta^8 \right]  )
+ \delta_B ( \left[ v \theta^6 \right] ) 
+  v^i \frac{\partial}
{\partial \phi^i } \left( \epsilon L^{(6)} \theta
 \right) = 0 ~ ,
\label{sevenf}
\end{equation}
modulo acceleration terms and higher fermion derivative terms.  Modulo
the same terms, the time derivative $d/ d \lambda$ has been replaced
as
\begin{equation}
 \frac{d}{d \lambda} \rightarrow v^i \frac{\partial}
{\partial \phi^i } ~.
\end{equation}
We also absorbed the possible total derivative terms (if any) into
$L^{(p)}$.  In general, considering the fact that $\theta$ is
multiplied from the right side of $L^{(p)}$ in
Eqs.~(\ref{onef})-(\ref{sevenf}), the part of matrices $L^{(p)}$ that
can give non-trivial contributions to Eqs.~(\ref{onef})-(\ref{sevenf})
can be expanded as
\begin{equation}
 L^{(p)} = a^{(p)} I_{16} + a^{(p)i} \gamma^i
 + a^{(p)ij} \gamma^{ij} + a^{(p)ijk} \gamma^{ijk}
 + a^{(p)ijkl} \gamma^{ijkl} ~ ,
\label{lexp}
\end{equation}
where $a^{(p)}$'s are $SO(9)$ totally anti-symmetric tensors made of
$p$-fermions $\theta$, $(3-p/2)$-vectors $v^i$ and an appropriate
number of scalars $\phi^i$.  Each term in Eq.~(\ref{lexp}) has a
coefficient function depending only on an $SO(9)$ invariant $\phi$.
Our goal is to solve Eqs.~(\ref{onef})-(\ref{sevenf}) and
(\ref{ninef}) to determine the coefficient functions of the effective
action.

We first compute $\delta_F ( \left[ v^3 \theta^2 \right] )$
to get
\begin{equation}
\delta_F ( \left[ v^3 \theta^2 \right] ) =
2 f^{(2)}_1 ( v^2 )^2 \phi^i (\epsilon \gamma^i \theta )
-2  f^{(2)}_1 v^2 v^i 
 ( v^j \phi^j  ) (\epsilon \gamma^i \theta ) ~ ,
\label{ferm2}
\end{equation}
and we have
\begin{equation}
\delta_B ( \left[ v^4 \right] ) = 
- i \left( \frac{d}{\phi d \phi } \right) f^{(0)} 
 (v^2 )^2 \phi^i  (\epsilon \gamma^i \theta ) ~. 
\label{boson0}
\end{equation}
We plug Eqs.~(\ref{ferm2}) and (\ref{boson0}) into Eq.~(\ref{onef}).
We note that the two terms of Eq.~(\ref{ferm2}) can not cancel with
each other.  Since Eqs.~(\ref{ferm2}) and (\ref{boson0}) contain only
$(\epsilon \gamma^i \theta )$, we can set $a^{(0)}$, $a^{(0)ij}$,
$a^{(0)ijk}$ and $a^{(0)ijkl}$ to zero in Eq.~(\ref{lexp}).
Furthermore, the possible terms of $a^{(0)i}$ can not contain factors
like $(v^i \phi^i)^n$ ($n > 0$), for the partial derivative $v^i
\partial / ( \partial \phi^i) $ then produces $(v^i \phi^i )^{n+1}$
terms when acting on their coefficient functions.  The resulting terms
are not present in Eqs.~(\ref{ferm2}) and (\ref{boson0}).  This leaves
us with a unique possibility
\begin{equation}
 \epsilon L^{(0)} \theta
 = h^{(0)} v^2 v^i ( \epsilon \gamma^i \theta ) . 
\label{l0}
\end{equation}
Upon inserting Eq.~(\ref{l0}) into Eq.~(\ref{onef}), the second term
of Eq.~(\ref{ferm2}) should cancel the term from Eq.~(\ref{l0})
resulting
\begin{equation}
\left( \frac{d}{\phi d \phi} \right) 
h^{(0)} = 2 f^{(2)}_1 . 
\end{equation}  
The first term of 
Eq.~(\ref{ferm2}) should cancel Eq.~(\ref{boson0})
to yield
\begin{equation}
 f^{(2)}_1 = \frac{i}{2} \left( \frac{d}{\phi d \phi}
\right) f^{(0)} ~ .
\label{f21}
\end{equation}
The spin-orbit coupling term $f^{(2)}_1$ is now determined in terms of
the bosonic coefficient function $f^{(0)}$.  It is identical to the
one-loop result computed in Ref.~\cite{kraus2} using the perturbative
matrix theory framework.

Going to Eq.~(\ref{threef}),
we compute
\begin{eqnarray}
 \delta_F ( \left[ v^2 \theta^4 \right] ) &=&
	4 v^2 ( f^{(4)}_2  v^i \phi^j \phi^k 
	+ f^{(4)}_0 v^i \delta^{jk} ) 
	  (\epsilon \gamma^k \theta )
	  ( \theta \gamma^{ij} \theta) \nonumber \\
    & &  - 4 f^{(4)}_2 (v^l \phi^l ) v^i v^k \phi^j 
	   (\epsilon \gamma^k \theta )
	( \theta \gamma^{ij}  \theta ) \nonumber \\
    & &  + 2 g^{(4)}_2 v^2 
	  \left( \phi^j \phi^k v^i
		(\epsilon \gamma^{ijl} \theta ) 
		(\theta \gamma^{kl} \theta ) 
		- \phi^i \phi^k v^j 
			(\epsilon \gamma^i \theta )
			(\theta \gamma^{kj} \theta )
	  \right. 		       \nonumber \\
    & &   ~~~~~~~~~~~~~
	  \left. + (v^k \phi^k ) \phi^i 
		(\epsilon \gamma^j \theta ) 
		(\theta \gamma^{ij} \theta ) 
	  \right) ~ , \label{ferm4} 
\end{eqnarray}
and
\begin{equation}
\delta_B ( \left[ v^3  \theta^2 \right]   ) =
-i v^2 \left( \left( \frac{d}{\phi d \phi} \right) f^{(2)}_1  
  v^i \phi^j \phi^k 
+ f^{(2)}_1 v^i \delta^{jk} \right) 
(\epsilon \gamma^k \theta )
 (\theta \gamma^{ij} \theta ) .
\label{boson2}
\end{equation}
We note that the first term of the $g^{(4)}_2$-dependent terms of
Eq.~(\ref{ferm4}) can not be canceled with any other terms of
Eqs.~(\ref{ferm4}) and (\ref{boson2}).  For the same reason as before,
we set $a^{(2)} = 0$ and consider terms of $\epsilon L^{(2)} \theta$
that do not contain $(v^i \phi^i )^n$ ($n > 0$) terms.  All possible
candidates from $\epsilon L^{(2)} \theta$ are as follows:
\begin{equation}
(\epsilon \gamma^{i} \theta ) (\theta
\gamma^{ij} \theta ) v^2 \phi^j 
\label{l21}
\end{equation}
\begin{equation}
(\epsilon \gamma^{i} \theta ) (\theta
\gamma^{jk} \theta ) v^i  v^j \phi^k
\label{l22}
\end{equation}
\begin{equation}
(\epsilon \gamma^{ij} \theta ) (\theta
\gamma^{kl} \theta ) v^i \phi^j v^k \phi^l
\label{l23}
\end{equation}
\begin{equation}
(\epsilon \gamma^{ij} \theta ) (\theta
\gamma^{jk} \theta ) v^i v^k 
\label{l24}
\end{equation}
\begin{equation}
(\epsilon \gamma^{ij} \theta ) (\theta
\gamma^{jk} \theta ) v^2 \phi^i \phi^k
\label{l25}
\end{equation}
\begin{equation}
(\epsilon \gamma^{ij} \theta ) (\theta
\gamma^{jkl} \theta ) v^i v^k \phi^l
\label{l26}
\end{equation}
\begin{equation}
(\epsilon \gamma^{ijk} \theta ) (\theta
\gamma^{jkl} \theta ) v^2 \phi^i \phi^l
\label{l27}
\end{equation}
\begin{equation}
(\epsilon \gamma^{ijk} \theta ) (\theta
\gamma^{jkl} \theta ) v^i v^l
\label{l28}
\end{equation}
\begin{equation}
(\epsilon \gamma^{ijk} \theta ) (\theta
\gamma^{kl} \theta ) v^i \phi^j v^l
\label{l29}
\end{equation}
\begin{equation}
(\epsilon \gamma^{ijk} \theta ) (\theta
\gamma^{klm} \theta ) v^i \phi^j v^l \phi^m
\label{l210}
\end{equation}
\begin{equation}
(\epsilon \gamma^{ijkl} \theta ) (\theta
\gamma^{klm} \theta ) \phi^i v^j v^m
\label{l211}
\end{equation}
recalling the Fierz identities in Appendix B.  The terms of
Eqs.~(\ref{l23})-(\ref{l211}) look as if they can possibly cancel the
$g^{(4)}_2$-dependent terms of Eq.~(\ref{ferm4}).  However, once the
derivative $v^i \partial / \partial \phi^i $ is taken, the maximum
scalar number of the terms resulting from
Eqs.~(\ref{l23})-(\ref{l211}) that do not contain the $(v^i \phi^i )$
factor is one, while the $g^{(4)}_2$-dependent terms in
Eq.~(\ref{ferm4}) has the maximum scalar number of two.  Therefore,
there are no other terms to cancel the $g^{(4)}_2$-dependent terms in
Eq.~(\ref{threef}) and this gives a non-trivial result
\begin{equation}
g^{(4)}_2 = 0 .
\label{g42}
\end{equation}
At the same time, we are now forced to set all the terms of
Eqs.~(\ref{l23})-(\ref{l211}) to zero. From the perturbative one-loop
four fermion terms calculated in Ref.~\cite{mcart}, we know that,
perturbatively, the spin-spin terms are absent among the four fermion
terms.  Eq.~(\ref{g42}) is the non-perturbative version of the same
statement.  To solve the remaining equations, we set
\begin{equation}
 \epsilon L^{(2)} \theta 
= \tilde{h}^{(2)}_1 v^2 \phi^j (\epsilon \gamma^i \theta )
 (\theta \gamma^{ij} \theta ) 
 +  h^{(2)}_1 v^i v^j \phi^k (\epsilon \gamma^i \theta )
 (\theta \gamma^{jk} \theta ) , 
\label{l20}
\end{equation}
from Eqs.~(\ref{l21}) and (\ref{l22}), which yields
\begin{eqnarray}
& & ( v^k \phi^k ) \left( \frac{d}{\phi d \phi} \right)
    \tilde{h}^{(2)}_1 v^2 \phi^j (\epsilon \gamma^i \theta )
    (\theta \gamma^{ij} \theta )  +
    ( v^l \phi^l )\left( \frac{d}{\phi d \phi} \right)
    h^{(2)}_1 v^i v^j \phi^k (\epsilon \gamma^i \theta )
    (\theta \gamma^{jk} \theta )
			\nonumber \\
& & + \tilde{h}^{(2)}_1  v^2 v^j (\epsilon \gamma^i \theta )
 (\theta \gamma^{ij} \theta ) ~ , 
\label{whew}
\end{eqnarray}
upon the differentiation in Eq.~(\ref{threef}). From 
Eqs.~(\ref{ferm4}), (\ref{boson2}), (\ref{g42})
and (\ref{whew}), we immediately find
\begin{equation}
 \tilde{h}^{(2)}_1 = 0 ~ , ~ \left( \frac{d}{\phi d \phi } 
 \right)
h^{(2)}_1 =   4 f^{(4)}_2 ~ ,
\end{equation}
\begin{equation}
 f^{(4)}_2 = \frac{i}{4} \left( \frac{d}{\phi d \phi} 
 \right) f^{(2)}_1 = - \frac{1}{8} 
  \left( \frac{d}{\phi d \phi} 
 \right)^2   f^{(0)}   ~ , 
\label{f42}
\end{equation}
and
\begin{equation}
 f^{(4)}_0 = \frac{i}{4} f^{(2)}_1 = - \frac{1}{8} 
 \left( \frac{d}{\phi d \phi} 
 \right)  f^{(0)} ~ .
\label{f40}
\end{equation}
These are precisely the same as the perturbatively calculated
one-loop spin-orbit four fermion terms \cite{mcart}.

We now consider Eq.~(\ref{fivef}).  We compute
\begin{eqnarray}
\delta_F ( \left[ v \theta^6 \right] ) & = &
f^{(6)}_3 v^i \phi^j \phi^k \phi^l 
\delta_F \left[ 
(\theta \gamma^{ij} \theta )
( \theta \gamma^{km} \theta )
( \theta \gamma^{lm} \theta ) \right] \nonumber \\
 & & + 2 f^{(6)}_1 v^i \phi^k
\delta_F \left[
(\theta \gamma^{ij} \theta )
( \theta \gamma^{jm} \theta )
( \theta \gamma^{km} \theta ) \right] ~ ,
\label{ferm6}
\end{eqnarray}
and
\begin{eqnarray}
\delta_B ( \left[ v^2 \theta^4 \right] ) & = &
 -i \left( \frac{d}{\phi d \phi} \right) f^{(4)}_2
 v^i v^p \phi^j  \phi^k  \phi^l 
(\epsilon \gamma^l \theta )
( \theta \gamma^{ij} \theta )
( \theta \gamma^{pk} \theta ) \nonumber \\
 & & -i f^{(4)}_2 v^i v^p \phi^k \left[ 
(\epsilon \gamma^k \theta )
( \theta \gamma^{ij} \theta )
( \theta \gamma^{pj} \theta ) 
+2 (\epsilon \gamma^j \theta )
( \theta \gamma^{ij} \theta )
( \theta \gamma^{pk} \theta ) \right] ~ ,
\label{boson4}
\end{eqnarray}
where the (apparent) three scalar term of Eq.~(\ref{ferm6}) 
is given by
\begin{eqnarray}
\lefteqn{ v^i \phi^j \phi^k \phi^l 
	\delta_F \left[ 
	(\theta \gamma^{ij} \theta )
	(\theta \gamma^{km} \theta )
	(\theta \gamma^{lm} \theta ) \right] = 
} \hspace{15mm}
			\nonumber \\
& &  2 v^i \phi^j \phi^k 
     \phi^l 
     \Big[ \:
	    v^i	(\epsilon \gamma^j \theta )
		( \theta \gamma^{km} \theta )
		( \theta \gamma^{lm} \theta )  
	  - v^j	(\epsilon \gamma^i \theta )
		( \theta \gamma^{km} \theta )
		( \theta \gamma^{lm} \theta ) 
			\nonumber \\
& &  \hspace{19mm}
	+2 v^p (\epsilon \gamma^{mpk} \theta )
		( \theta \gamma^{lm} \theta )
		( \theta \gamma^{ij} \theta )
	+2 v^k 	(\epsilon \gamma^m \theta )
		( \theta \gamma^{lm} \theta )
		( \theta \gamma^{ij} \theta )
			\nonumber \\
& &  \hspace{19mm}	
	-2 v^m 	(\epsilon \gamma^k \theta )
		( \theta \gamma^{lm} \theta )
		( \theta \gamma^{ij} \theta ) \:
     \Big]  ~ ,
\label{fake3}
\end{eqnarray}
and the (apparent) one scalar term of Eq.~(\ref{ferm6}) is 
\begin{eqnarray}
\lefteqn{ v^i \phi^k
	\delta_F \left[
			(\theta \gamma^{ij} \theta )
			( \theta \gamma^{jm} \theta )
			( \theta \gamma^{km} \theta ) 
		 \right] = 
} \hspace{15mm}
			\nonumber \\
& &  2 \phi^k 
     \Big[ \:
	    v^2	(\epsilon \gamma^j \theta )
	    	( \theta \gamma^{jm} \theta )
		( \theta \gamma^{km} \theta ) 
      - v^i v^p (\epsilon \gamma^i \theta )
		( \theta \gamma^{pm} \theta )
		( \theta \gamma^{km} \theta )
			\nonumber \\
& &   \hspace{9mm}
      + v^i v^p (\epsilon \gamma^{pkm} \theta )
		( \theta \gamma^{jm} \theta )
		( \theta \gamma^{ij} \theta ) 
      + v^i v^p (\epsilon \gamma^{pjm} \theta )
		( \theta \gamma^{km} \theta )
		( \theta \gamma^{ij} \theta )
			\nonumber \\
& &   \hspace{9mm}
      + v^i v^k (\epsilon \gamma^m \theta )
		( \theta \gamma^{jm} \theta )
		( \theta \gamma^{ij} \theta ) 
      - v^i v^p (\epsilon \gamma^k \theta )
		( \theta \gamma^{jp} \theta )
		( \theta \gamma^{ij} \theta )
			\nonumber \\
& &   \hspace{9mm}
      - v^i v^p (\epsilon \gamma^j \theta )
		( \theta \gamma^{kp} \theta )
		( \theta \gamma^{ij} \theta ) 
     \: \Big] ~ .
\label{fake1}
\end{eqnarray}
Among the terms of Eqs.~(\ref{fake3}) and (\ref{fake1}), there are
terms with a single $(v^i \phi^i)$ factor and terms of the form
$(\epsilon \gamma^{mpk} \theta ) ( \theta \gamma^{lm} \theta ) (
\theta \gamma^{ij} \theta )$.  Via Fierz identities, the second group
of terms can be reduced to the simpler terms like the ones in
Eq.~(\ref{boson4}).  In this process, some of the indices appearing in
the fermion bilinears are contracted, resulting the contractions among
the $v$'s and $\phi$'s.  Closer inspection of the Fierz identities and
the structure of the terms of Eqs.~(\ref{fake3}) and (\ref{fake1})
show that the double contractions of $v$'s and $\phi$'s vanish and
only a single contraction is allowed for the terms of
Eqs.~(\ref{fake3}) and (\ref{fake1}).  Therefore, the terms of
Eqs.~(\ref{ferm6}) and (\ref{boson4}) are classified as shown in Table
2.  
\\
\begin{center}
\begin{tabular}{|c|c|c|c|c|c|c|} \hline
& $(r,s)$  &  (\ref{boson4}) & (\ref{fake3}) & (\ref{fake1})
        &  $h^{(4)}_2$    & $h^{(4)}_0$    \\ \hline 
$(a)$ & (3,0)  &  $\left( \frac{d}{\phi d \phi} \right) f^{(4)}_2$
        &  $f^{(6)}_3$    &   &   &  \\ \hline
$(b)$ &  (2,1)  &  &  $(v^i \phi^i) f^{(6)}_3 $  & 
        &  $ (v^i \phi^i ) \left( \frac{d}{\phi d \phi} \right)
           h^{(4)}_2 $  &  \\ \hline
$(c)$ &  (1,0)  &  $f^{(4)}_2 $ & $\phi^2 f^{(6)}_3 $ & $f^{(6)}_1$
        &  $h^{(4)}_2 $ &                \\ \hline
$(d)$ & (0,1)  &  &  & $(v^i \phi^i) f^{(6)}_1 $ 
        &                 &    $ (v^i \phi^i ) 
\left( \frac{d}{\phi d \phi} \right) h^{(4)}_0 $  \\ \hline
\end{tabular}
\\
[.3cm]
\end{center}
{\small Table 2.  The classification of the terms appearing in
Eq.~(\ref{fivef}).  Pairs of numbers $(r,s)$ shown in the second
column denote the scalar number and the number of $(v^i \phi^i)$,
respectively.  The terms with different $(r,s)$ can not cancel with each other.
Each entry in the table shows the $SO(9)$ scalar factor of each term,
suppressing the degeneracy and the tensor structures. }  
\\ [.3cm]
Also shown in Table 2 is the classification of the terms from
$\epsilon L^{(4)} \theta$ that show up in Eq.~(\ref{fivef}).
Generally, the possible terms of $\epsilon L^{(4)} \theta$ include
zero, one, two and three scalar structure terms whose scalar
coefficient functions we denote as $h^{(4)}_q$ where $q$ ($q=0 ,
1,2,3$) is the number of scalars, since $\epsilon L^{(4)} \theta
\propto (\epsilon \gamma^{i_1 \cdots} \theta ) (\theta \gamma^{j_1
\cdots} \theta ) (\theta \gamma^{k_1 \cdots} \theta )$.  The three
scalar structure terms of $\epsilon L^{(4)} \theta$ produce $(3,1)$
and $(2,0)$ terms that do not appear elsewhere in Eq.~(\ref{fivef}),
upon taking the derivative $v^i \partial / ( \partial \phi^i )$.
Likewise, the one scalar structure terms of $\epsilon L^{(4)} \theta$
yield $(1,1)$ and $(0,0)$ terms in Eq.~(\ref{fivef}), which are also
absent in Table 2.  As such, the coefficient functions $h^{(4)}_3 =
h^{(4)}_1 = 0$.  Therefore, only the two scalar and zero scalar terms
of $\epsilon L^{(4)} \theta$ are non-vanishing and they correspond to
the $h^{(4)}_2$ and $h^{(4)}_0$ columns of Table 2.

It is seen to be clear how to determine the scalar coefficient
functions.  We notice from the row $(a)$ of Table 2 that there are
no contributions from $\epsilon L^{(4)} \theta$ for the maximum
scalar structure terms of type $(3,0)$\footnote{Precisely this
observation was used to simplify the membrane spin-orbit coupling
calculations of Ref.~\cite{membrane}.}.  The $(3,0)$ terms from
Eq.~(\ref{fake3}) should directly cancel the maximum scalar number
terms of Eq.~(\ref{boson4}), yielding $f^{(6)}_3 \propto (d / (\phi d
\phi )) f^{(4)}_2$.  Next, from the row $(b)$ of Table 2, $h^{(4)}_2$
is determined to give $h^{(4)}_2 \propto f^{(4)}_2$ via $(d / (\phi d
\phi )) h^{(4)}_2 \propto f^{(6)}_3$.  The third row $(c)$ determines
$f^{(6)}_1$ in terms of $f^{(4)}_2$, $\phi^2 f^{(6)}_3$ and
$h^{(4)}_2$ to be $f^{(6)}_1 \propto f^{(4)}_2$.  Finally, from the
row $(d)$, the functions $h^{(4)}_0$ are obtained as $h^{(4)}_0
\propto f^{(4)}_0$.  Hereafter, we present the detailed derivation of
$f^{(6)}_3$ from the row $(a)$.  One technical comment should be in
order; as the number of fermions increases, we need progressively more
complicated Fierz identities.  Especially when the Fierz identities
involve two different constant spinors $\epsilon$ and $\theta$, they
become even more complicated.  For the simplification of the
computations, we note that an arbitrary $\epsilon$ can be obtained by
multiplying $\theta$ with an appropriate $16 \times 16$ matrix.
Recalling a complete expansion of the form (\ref{lexp}), it is thus
equivalent to consider $\theta$ (which typically produces trivial
results), $\theta \gamma^i$, $\theta \gamma^{ij}$, $\theta
\gamma^{ijk}$ and $\theta \gamma^{ijkl}$ in place of $\epsilon$.  From
now on, we will replace $\epsilon$ with $\theta \gamma^i$.  The cases
for $\theta \gamma^{ij}$, $\theta \gamma^{ijk}$ and $\theta
\gamma^{ijkl}$ can be analyzed in a similar fashion to show the
complete consistency.

Upon replacing $\epsilon \rightarrow \theta \gamma^n$ and retaining
the maximum three scalar terms of Eq.~(\ref{boson4}), we have
\begin{equation}
 -i \left( \frac{d}{\phi d \phi} \right) f^{(4)}_2
 v^i v^p \phi^j \phi^k \phi^l 
 (\theta \gamma^{nl} \theta )
 (\theta \gamma^{ij} \theta )
 (\theta \gamma^{pk} \theta ) ~,
\label{eee}
\end{equation}
and
\begin{eqnarray}
& & 2 f^{(6)}_3 v^i v^p \phi^j \phi^k \phi^l 
  \left[ \delta^{ip} 
   (\theta \gamma^{nj} \theta )
 (\theta \gamma^{km} \theta )
 (\theta \gamma^{lm} \theta )  +2  \delta^{np} 
   (\theta \gamma^{ij} \theta )
 (\theta \gamma^{km} \theta )
 (\theta \gamma^{lm} \theta ) \right. \nonumber \\
& & \hspace{30mm} \left. -2  \delta^{nk} 
   (\theta \gamma^{ij} \theta )
 (\theta \gamma^{pm} \theta )
 (\theta \gamma^{lm} \theta ) \right] ~,
\label{eeee}
\end{eqnarray}
from Eq.~(\ref{fake3}) (apparent) three scalar terms using the Wick
theorem.  We have to check whether the terms of Eq.~(\ref{eeee}) turn
into the terms of Eq.~(\ref{eee}) up to $(v^i \phi^i )$ terms and the
lower scalar number terms.  Using the Fierz identities
Eqs.~(\ref{fier4})-(\ref{fier6}), we can show that:
\begin{eqnarray}
\lefteqn{ v^i v^p \phi^j \phi^k \phi^l 
   \left[ \delta^{ip} 
   (\theta \gamma^{nj} \theta )
 (\theta \gamma^{km} \theta )
 (\theta \gamma^{lm} \theta )  +2  \delta^{np} 
   (\theta \gamma^{ij} \theta )
 (\theta \gamma^{km} \theta )
 (\theta \gamma^{lm} \theta )   
  \right.} 	\hspace{50mm}	\nonumber \\
\lefteqn{ \left. -2  \delta^{nk} 
   (\theta \gamma^{ij} \theta )
 (\theta \gamma^{pm} \theta )
 (\theta \gamma^{lm} \theta ) \right] }
		\hspace{30mm}	 \nonumber \\
&=&  v^i v^p \phi^j \phi^k \phi^l 
	\left[
	 45 (\theta \gamma^{nl} \theta )
 	    (\theta \gamma^{ij} \theta )
 	    (\theta \gamma^{pk} \theta )
	\right.		\nonumber \\
& & \hspace{22mm} -4 \delta^{ip} 
   	(\theta \gamma^{nj} \theta )
 	(\theta \gamma^{km} \theta )
 	(\theta \gamma^{lm} \theta )
			\nonumber \\  
& & \hspace{22mm} -8  \delta^{np} 
   	(\theta \gamma^{ij} \theta )
 	(\theta \gamma^{km} \theta )
 	(\theta \gamma^{lm} \theta )
			\nonumber \\
& & \hspace{22mm} \left. +8 \delta^{nk} 
   	(\theta \gamma^{ij} \theta )
 	(\theta \gamma^{pm} \theta )
 	(\theta \gamma^{lm} \theta )
     \right]		\nonumber \\
& &  + (v^i \phi^i) {\rm ~terms~and~
       lower~scalar~terms} ~.
\label{monster1}
\end{eqnarray}
We start from using Fierz identity Eq.~(\ref{fier3}).  We sequentially
use the Fierz identity Eq.~(\ref{fier4}) and the five free index Fierz
identities, Eqs.~(\ref{fierfor6})-(\ref{fier6}) to reduce the terms of
the left hand side into the form where we can use the Fierz identity
Eq.~(\ref{fier3}).  Then, using Eq.~(\ref{fier3}) again further
reduces all the terms into the three products of $(\theta \gamma^{ij}
\theta )$'s, producing the identity Eq.~(\ref{monster1}).  In this
process, noticing the permutation symmetry of $\phi^j \phi^k \phi^l$
simplifies the calculations.  Eq.~(\ref{monster1}) can be rearranged
to yield:
\begin{eqnarray}
\lefteqn{ v^i v^p \phi^j \phi^k \phi^l 
   \left[ \delta^{ip} 
   (\theta \gamma^{nj} \theta )
 (\theta \gamma^{km} \theta )
 (\theta \gamma^{lm} \theta )  +2  \delta^{np} 
   (\theta \gamma^{ij} \theta )
 (\theta \gamma^{km} \theta )
 (\theta \gamma^{lm} \theta )   
  \right.} 	\nonumber \\
\lefteqn{ \hspace{22mm}
	\left. -2  \delta^{nk} 
   (\theta \gamma^{ij} \theta )
 (\theta \gamma^{pm} \theta )
 (\theta \gamma^{lm} \theta ) \right] }
			 \nonumber \\
&=&  9 v^i v^p \phi^j \phi^k \phi^l
	    (\theta \gamma^{nl} \theta )
 	    (\theta \gamma^{ij} \theta )
 	    (\theta \gamma^{pk} \theta )
   + (v^i \phi^i) {\rm ~terms~and~
       lower~scalar~terms} ~.
\label{hihihi}
\end{eqnarray}
Now Eq.~({\ref{hihihi}) immediately implies that Eq.~(\ref{eeee})
becomes 
\begin{equation}
18 f^{(6)}_3
 v^i v^p \phi^j \phi^k \phi^l 
 (\theta \gamma^{nl} \theta )
 (\theta \gamma^{ij} \theta )
 (\theta \gamma^{pk} \theta )
+ (v^i \phi^i) {\rm ~terms~and~
       lower~scalar~terms} ~,
\end{equation}
and, via Eq.~(\ref{fivef}), yields
\begin{equation}
f^{(6)}_3 =  \frac{i}{18} \left( \frac{d}{\phi d \phi}
\right) f^{(4)}_2
= -\frac{i}{144} \left( \frac{d}{\phi d \phi}
\right)^3 f^{(0)} .
\label{f63} 
\end{equation}
Now that we determined $f^{(6)}_3$, we can recursively solve the row
$(b)$ and $(c)$ of Table 2 to determine the function $f^{(6)}_1$.  The
result is:
\begin{equation}
f^{(6)}_1 = \frac{i}{18}  f^{(4)}_2
= -\frac{i}{144} \left( \frac{d}{\phi d \phi}
\right)^2 f^{(0)} .
\label{f61}
\end{equation}
We remark that we again need to use the five free index Fierz
identities shown in Appendix B.  The terms of the effective action,
Eqs.~(\ref{f63}) and (\ref{f61}), for six fermion mixed spin-orbit
spin-spin couplings have not yet been calculated within the
perturbative matrix theory framework.  However, Eqs.~(\ref{f63}) and
(\ref{f61}) are identical to the corresponding terms computed from the
perturbative IIA string theory framework \cite{morales,review}.

Before proceeding to Eq.~(\ref{sevenf}), we recall that the eight
fermion terms in the effective action can be completely determined up
to an overall normalization by solving Eq.~(\ref{ninef})
\cite{sethi2}.  Thus, the remaining task is only to consider the
maximum scalar number terms of Eq.~(\ref{sevenf}), which links the
bosonic variation of the three scalar term of the six fermion terms to
the fermionic variation of the four scalar term of the eight fermion
terms.  As noted in the computation of the six fermion terms, the
knowledge of $\epsilon L^{(6)} \theta$ terms is not necessary for this
purpose.  Furthermore, replacing $\epsilon$ with $\theta \gamma^i$ is
enough to get the desired result.  The consideration of $\theta$,
$\theta \gamma^{ij}$, $\theta \gamma^{ijk}$ and $\theta \gamma^{ijkl}$
in place of $\epsilon$ can be straightforwardly performed to show the
complete consistency, although the computations are quite lengthy in
these cases.  We compute the fermionic variation of the eight fermion
terms
\begin{eqnarray}
\delta_F ( \left[ \theta^8 \right] ) &=&
   f^{(8)}_4 \phi^i \phi^j \phi^k \phi^l 
	\delta_F \left[ (\theta \gamma^{im} \theta)
			(\theta \gamma^{jm} \theta)
			(\theta \gamma^{kn} \theta)
			(\theta \gamma^{ln} \theta)
		\right]  
			\nonumber \\
  & & +4 f^{(8)}_2 \delta^{ik} \phi^j  \phi^l 
	\delta_F \left[ (\theta \gamma^{im} \theta)
			(\theta \gamma^{jm} \theta)
			(\theta \gamma^{kn} \theta)
			(\theta \gamma^{ln} \theta)
		\right]
			\nonumber \\
  & & +2f^{(8)}_0 \delta^{ik} \delta^{jl} 
	\delta_F \left[ (\theta \gamma^{im} \theta)
			(\theta \gamma^{jm} \theta)
			(\theta \gamma^{kn} \theta)
			(\theta \gamma^{ln} \theta)
		\right]  \label{ferm8}
\end{eqnarray}
and the bosonic variation of the three scalar term
of the six fermion terms
\begin{eqnarray}
\delta_B ( \left[ v \theta^6 \right] ) &=&
	-i \left( \frac{d}{\phi d \phi} \right) f^{(6)}_3
		v^p \phi^i \phi^j \phi^k \phi^l
     			(\epsilon \gamma^i \theta)
			(\theta \gamma^{pj} \theta)
			(\theta \gamma^{kn} \theta)
			(\theta \gamma^{ln} \theta)
				\nonumber \\
 & & + {\rm ~ lower~ scalar~ terms} ~,
\label{boson6}
\end{eqnarray}
both of which appear in Eq.~(\ref{sevenf}).
Written explicitly, the (apparent) four scalar term
of Eq.~(\ref{ferm8}) is given by     
\begin{eqnarray}
\lefteqn{ f^{(8)}_4 \phi^i \phi^j \phi^k \phi^l 
	 \delta_F \left[ (\theta \gamma^{im} \theta)
			(\theta \gamma^{jm} \theta)
			(\theta \gamma^{kn} \theta)
			(\theta \gamma^{ln} \theta)
		 \right] = }
			\nonumber \\
 & & 8 f^{(8)}_4 \phi^i \phi^j \phi^k \phi^l 
	\left[
            v^p (\epsilon \gamma^{pim} \theta)
 		(\theta \gamma^{jm} \theta)
		(\theta \gamma^{kn} \theta)
		(\theta \gamma^{ln} \theta)
	  - v^m (\epsilon \gamma^{i} \theta)
		(\theta \gamma^{jm} \theta)
		(\theta \gamma^{kn} \theta)
		(\theta \gamma^{ln} \theta)
	\right] ~ . 
			\nonumber \\
 & &
\label{maxferm8}
\end{eqnarray}
Upon the replacement $ \epsilon \rightarrow \theta \gamma^{s} $, the
four scalar terms of Eq.~(\ref{boson6}) become
\begin{equation}
-i \left( \frac{d}{\phi d \phi} \right) f^{(6)}_3
 \phi^i \phi^j \phi^k \phi^l v^p 
	        (\theta \gamma^{si} \theta)
	 	(\theta \gamma^{pj} \theta)
		(\theta \gamma^{kn} \theta)
		(\theta \gamma^{ln} \theta) ~ ,
\label{fff}
\end{equation}
and Eq.~(\ref{maxferm8}) becomes
\begin{equation}
 8 f^{(8)}_4 \phi^i \phi^j \phi^k \phi^l v^p 
	\left[
            \delta^{si} (\theta \gamma^{pm} \theta)
 		(\theta \gamma^{mj} \theta)
		(\theta \gamma^{kn} \theta)
		(\theta \gamma^{ln} \theta)
	  + \delta^{sp} (\theta \gamma^{mi} \theta)
		(\theta \gamma^{mj} \theta)
		(\theta \gamma^{kn} \theta)
		(\theta \gamma^{ln} \theta)
	\right]  ~,
\label{ffff}
\end{equation}
using the Wick theorem.  We have to check whether the 
terms of Eq.~(\ref{ffff})
become the form of Eq.~(\ref{fff}) 
up to $(v^i \phi^i )$ terms and the lower
scalar number terms.  Using the Fierz identities
in Appendix B, we can show the
following identity:
\begin{eqnarray}
\lefteqn{
  \phi^i \phi^j \phi^k \phi^l v^p 
	\left[
            \delta^{si} (\theta \gamma^{pm} \theta)
 		(\theta \gamma^{mj} \theta)
		(\theta \gamma^{kn} \theta)
		(\theta \gamma^{ln} \theta)
	  + \delta^{sp} (\theta \gamma^{mi} \theta)
		(\theta \gamma^{mj} \theta)
		(\theta \gamma^{kn} \theta)
		(\theta \gamma^{ln} \theta)
	\right]
}   \hspace{40mm}
			\nonumber \\
 &=& \phi^i \phi^j \phi^k \phi^l v^p 
     \left[
	    -35 (\theta \gamma^{si} \theta)
	 	(\theta \gamma^{pj} \theta)
		(\theta \gamma^{kn} \theta)
		(\theta \gamma^{ln} \theta)
     \right.
			\nonumber \\
 & &	  \hspace{22mm}   
	     +6 \delta^{si}
		(\theta \gamma^{pm} \theta)
		(\theta \gamma^{mj} \theta)
		(\theta \gamma^{kn} \theta)
		(\theta \gamma^{ln} \theta)
			\nonumber \\
 & & 	  \hspace{22mm} 
     \left.
	     +6 \delta^{sp} 
		(\theta \gamma^{mi} \theta)
		(\theta \gamma^{mj} \theta)
		(\theta \gamma^{kn} \theta)
		(\theta \gamma^{ln} \theta)
     \right]  \nonumber \\
 & &   +  (v^i \phi^i) {\rm ~terms~and~
  lower~scalar~terms} ~.
\label{monster2}
\end{eqnarray}
In deriving Eq.~(\ref{monster2}), we first use the Fierz identity
Eq.~(\ref{fier3}).  Next, Eq.~(\ref{fier4}) and the five free index
Fierz identities, Eqs.~(\ref{fier5}) and (\ref{fier6}), are utilized to
transform the terms on the left hand side into the form where we can
use the Fierz identity Eq.~(\ref{fier3}) again.  Then, upon using
Eq.~(\ref{fier3}), we reduce all the terms into four products of
$(\theta \gamma^{ij} \theta )$'s.  Using the identity
Eq.~(\ref{monster2}), we rewrite Eq.~(\ref{ffff}) as
\begin{eqnarray}
& & 56 f^{(8)}_4 \phi^i \phi^j \phi^k \phi^l v^p 
	        (\theta \gamma^{si} \theta)
	 	(\theta \gamma^{pj} \theta)
		(\theta \gamma^{kn} \theta)
		(\theta \gamma^{ln} \theta) \nonumber \\
& &   +  (v^i \phi^i) {\rm ~terms~and~
  lower~scalar~terms} ~,
\end{eqnarray}
which, via Eq.~(\ref{sevenf}), implies
\begin{equation}
f^{(8)}_4 = \frac{i}{56} \left( \frac{d}{\phi d \phi}
\right) f^{(6)}_3
= \frac{1}{8064} \left( \frac{d}{\phi d \phi}
\right)^4 f^{(0)} ~ .
\label{f84}
\end{equation}
By solving Eq.~(\ref{ninef}), the coefficient 
functions $f^{(8)}_4$, $f^{(8)}_2$ and 
$f^{(8)}_0$ are computed to be proportional to 
$\phi^{-15}$, $\phi^{-13}$ and $\phi^{-11}$, respectively,
with known relative coefficients \cite{sethi2}:
\begin{equation}
 f^{(8)}_4 = c \frac{1}{\phi^{15}} ~~ , ~~
 f^{(8)}_2 = - \frac{c}{13} \frac{1}{\phi^{13}} ~~ , ~~  
 f^{(8)}_0 =  \frac{c}{143} \frac{1}{\phi^{11}} ~ ,
\end{equation}
where $c$ is an overall constant.  We can immediately 
integrate Eq.~(\ref{f84}) to obtain
\begin{equation}
f^{(0)} = k_1 + k_2 \frac{1}{\phi^7} 
 + k_3 \phi^2 + k_4 \phi^4 + k_5 \phi^6 ~ ,
\label{finish}
\end{equation}
where $k_1$, $k_3$, $k_4$, $k_5$ are constants of
integration and $k_2 = (128 / 143 )  c$.  
The $\phi \rightarrow \infty$ limit of the $F^4$
terms should be finite from the physical point of
view and we thus set $k_3 = k_4 = k_5 = 0$.
We therefore find that 
\begin{equation}
 f^{(8)}_2 = \frac{1}{8064} \left( \frac{d}{\phi d \phi} 
 \right)^3 f^{(0)} ~~ , ~~
 f^{(8)}_0 = \frac{1}{8064} \left( \frac{d}{\phi d \phi} 
  \right)^2 f^{(0)} ~ .
\label{f820}
\end{equation}
Eqs.~(\ref{f84}) and (\ref{f820}) are identical to the
one-loop spin-spin terms calculated in Ref.~\cite{barrio}.
This completes our analysis
and we derived the results shown in Sec.~1, 
Eqs.~(\ref{first})-(\ref{last}),
by collecting Eqs. (\ref{v2act}), (\ref{finish}), (\ref{f21}), 
(\ref{g42}), (\ref{f42}), (\ref{f40}), (\ref{f63}), 
(\ref{f61}), (\ref{f84}) and (\ref{f820}).    

\section{Discussions}

The consideration in this paper has been restricted to the case of the
$M$-momentum scatterings.  However, for the extended objects, we can
also use the results of the analysis presented in this paper as far as
the `center of mass' dynamics is concerned.  For example, for the
membrane dynamics considered in Ref.~\cite{sethi} (see also
\cite{membrane}), the four scalar term of the eight fermion terms in
$n$-instanton sector has been computed to be
\begin{equation}
f^{(8)}_4 = k_1 n^{13/2} \phi^{-13/2} 
K_{13/2} ( n \phi /g) e^{i n \phi^8 /g} ,
\end{equation} 
and the two scalar term and the zero scalar term coefficient functions
$4f^{(8)}_2$ and $2 f^{(8)}_0$ are the inhomogeneous solutions of
\begin{equation}
\left( \frac{d^2}{d \phi^2 } + \frac{10}{\phi} 
  \frac{d}{d \phi} 
 - \frac{n^2}{g^4} \right) 4f^{(8)}_2 = - 8 f^{(8)}_4
\label{a1}
\end{equation}
and
\begin{equation}
 \left( \frac{d^2}{d \phi^2 } + \frac{6}{\phi} 
 \frac{d}{d \phi} 
 - \frac{n^2}{g^4} \right) 2f^{(8)}_0 = 
  - 2 (4 f^{(8)}_2 ) ~,
\label{a2}
\end{equation}
respectively.  Here $k_1$ and $g$ are dimensionful constants and
$K_{\nu}$ is the modified Bessel function with a half-integer
coefficient $\nu$.  The $SO(7)$ vectors $\phi^i$ ($i=1, \cdots , 7 $ )
combine to give an $SO(7)$ invariant $\phi^2 = \phi^i \phi^i$, and
$\phi^8$ is the dual magnetic scalar.  It can be easily shown that the
function $f_{\nu} = \phi^{- \nu} K_{\nu} (n \phi / g^2 )$ satisfies
the differential equation
\begin{equation}
\left(  \frac{d^2}{d \phi^2 } + \frac{2 \nu + 1}{\phi} 
  \frac{d}{d\phi } 
 - \frac{n^2}{g^4} \right) f_{\nu} = 0 ~ .
\label{a3}
\end{equation}
  From the recursion relation
\begin{equation}
 \left( \frac{d}{zdz} \right)^a ( z^{- \nu}  K_{\nu} (z)  )
= (-1)^a z^{- \nu - a} K_{\nu + a} (z)  , 
\label{wow}
\end{equation}
we can immediately write down the inhomogeneous solutions of
Eqs.~(\ref{a1}) and (\ref{a2}) as
\begin{equation}
 4 f^{(8)}_2 =  - 4 k_1 g n^{11/2}  
  \phi^{-11/2} K_{11/2} ( n \phi /g) e^{i n \phi^8 /g}
\label{a4}
\end{equation}
and
\begin{equation}
 2 f^{(8)}_0 =  2 k_1 g^2 n^{9/2} 
   \phi^{-9/2} K_{9/2} ( n \phi /g) e^{i n \phi^8 /g} ~,
\label{a5}
\end{equation}
where we simultaneously use Eq.~(\ref{a3}) to delete the second
derivative terms and the zero derivative terms of Eqs.~(\ref{a1}) and
(\ref{a2}).  Noting that
\begin{equation}
f^{(0)} = k_2 n^{5/2} \phi^{-5/2} K_{5/2} (n \phi / g )
 e^{i n \phi^8 / g } ~,
\end{equation}
where $k_2$ is a constant, from Ref.~\cite{membrane} and recalling
Eq.~(\ref{wow}), we conclude that Eqs.~(\ref{a4}) and (\ref{a5}) are
completely consistent with Eq.~(\ref{ansatz}) for
$p=4$ and $q = 0,2,4$.  It will be
interesting to apply this type of arguments to the higher brane
two-body dynamics.

The derivation presented in Sec.~2.2 does not appear to sensitively
depend on the existence of the unbroken $SO(9)$ $R$-symmetry, even if
the classification of the possible terms in Sec.~2.1 does.
Consequently, for an arbitrary point in the moduli space, that
generally breaks $SO(9)$ to its subgroup and represents an
arbitrarily separated source $M$-momenta, we 
write down the effective action (by the linear superposition 
of the source $M$-momenta, which is valid when there are
full supersymmetries):
\begin{eqnarray}
\Gamma_{(4)} &=&  
	\int d \lambda \Big[
	 f (v^2)^2 + \frac{i}{2} v^2 v^i \partial_j f 
  	(\theta \gamma^{ij} \theta ) 
	 - \frac{1}{8} v^i v^j \partial_k \partial_l f
  	(\theta \gamma^{ik} \theta ) 
  	(\theta \gamma^{jl} \theta ) 
			\nonumber \\
             & & ~~~~~~~~
	 - \frac{i}{144} v^i 
	\partial_j \partial_k \partial_l f
	(\theta \gamma^{ij} \theta)
	(\theta \gamma^{km} \theta ) 
  	(\theta \gamma^{lm} \theta ) 
			\nonumber \\
	     & & ~~~~~~~~
 	+ \frac{1}{8064} \partial_i \partial_j \partial_k 
 	\partial_l f (\theta \gamma^{im} \theta)
  	(\theta \gamma^{jm} \theta)
  	(\theta \gamma^{kn} \theta ) 
  	(\theta \gamma^{ln} \theta ) \Big]  
			\label{conjecture} ~,
\end{eqnarray}
where $f$ is an arbitrary nine-dimensional harmonic function that
vanishes as $\phi^i \rightarrow \infty$ and we use the normalization
convention of the quadratic terms of Eq.~(\ref{v2act}).  The formal
observation at the end of Sec.~2.1 is used to write down the action
(\ref{conjecture}).  Up to two fermion terms, Eq.~(\ref{conjecture})
agrees with the probe dynamics calculations in the DLCQ supergravity
framework using the multi-center $M$-momenta solutions as the
background geometry \cite{new}.  The multi-center background geometry
solutions of the DLCQ supergravity preserve the sixteen
supersymmetries as in the case of the single center solutions.
Furthermore, the BPS solution space for the $N$ source $M$-momenta
from the supergravity is the $N$-symmetric product of $R^9$, just like
the SYM quantum mechanics moduli space.

Beyond our analysis presented in this paper, a possible next step is
to repeat the same type of analysis to the $F^6$ terms.  The
supersymmetric completion of these terms, once determined, can be used
to prove the two-loop exactness of the $v^6$ term, which is a
necessary element in firmly establishing the matrix theory/DLCQ
supergravity correspondence.  Furthermore, at this order, we expect
that the matrix theory produces genuine quantum gravity corrections to
the eleven-dimensional supergravity.  It will be interesting to
explicitly compute these terms and compare them to the quantum
corrected DLCQ supergravity and to the type II stringy corrections.
Another very interesting issue, as mentioned in Sec.~1, is to
understand how much of the constructions presented here can survive
under the supersymmetry breaking.

\section*{Acknowledgments}
We would like to thank Sangmin Lee for useful discussions, and
Y.~K. would like to thank H.~Verlinde and S.~Sethi for stimulating
conversations. We are also grateful to W.~Taylor for bringing
Ref.~\cite{taylor} to our attention.

\appendix

\begin{center}
{\Large \bf Appendix}
\end{center}

\section{Reduction of 
$\theta \gamma^{ijk} \theta$ products
to $\theta \gamma^{ij} \theta$ products}

We start from writing all possible terms of $8^{v^2 KK}$.  There
are seven possible terms:
\[ (v^i \phi^i )^2 \phi^j \phi^k 
(\theta \gamma^{jlm} \theta ) 
(\theta \gamma^{klm} \theta ) ~ , \]
\[ v^2 \phi^i \phi^j 
(\theta \gamma^{ikl} \theta ) 
(\theta \gamma^{jkl} \theta ) ~ , \]
\[ v^i v^j \phi^k \phi^l
(\theta \gamma^{ikm} \theta ) 
(\theta \gamma^{jlm} \theta ) ~ , \]
\[ (v^i \phi^i ) v^j \phi^k
(\theta \gamma^{jlm} \theta ) 
(\theta \gamma^{klm} \theta ) ~ , \]
\[ (v^i \phi^i )^2
(\theta \gamma^{jkl} \theta ) 
(\theta \gamma^{jkl} \theta ) ~ , \]
\[ v^2
(\theta \gamma^{ijk} \theta ) 
(\theta \gamma^{ijk} \theta ) ~ , \]
\[ v^i v^j 
(\theta \gamma^{ikl} \theta ) 
(\theta \gamma^{jkl} \theta ) ~ . \]
Using the Fierz identities (\ref{fier1}),
(\ref{fier2}) and (\ref{fier3}),
we can show that all the terms in the above reduce to the terms of
$6^{v^2 JJ}$ or vanish.  In the case of the terms of $7^{v^2 JK}$,
there are terms like
\[ v^i v^j \phi^k 
(\theta \gamma^{il} \theta ) (\theta \gamma^{ljk} \theta ) , \] 
which can not be reduced to the terms of $6^{v^2 JJ}$ even if we use
Fierz identities; the number of scalars do not match.

The possible $9^{vJKK}$ terms are as follows.
\[ v^i \phi^j \phi^k \phi^l 
(\theta \gamma^{ij} \theta ) 
(\theta \gamma^{kmn} \theta )
(\theta \gamma^{lmn} \theta )  ~ , \]
\[ v^i \phi^j
(\theta \gamma^{ik} \theta ) 
(\theta \gamma^{kmn} \theta )
(\theta \gamma^{jmn} \theta )  ~ , \]
\[  v^i \phi^j 
(\theta \gamma^{ij} \theta ) 
(\theta \gamma^{klm} \theta )
(\theta \gamma^{klm} \theta )  ~ , \]
\[  v^i \phi^j \phi^k \phi^l
(\theta \gamma^{jm} \theta ) 
(\theta \gamma^{ikn} \theta )
(\theta \gamma^{lmn} \theta )  ~ , \]
\[  v^i \phi^m
(\theta \gamma^{jl} \theta ) 
(\theta \gamma^{ijk} \theta )
(\theta \gamma^{klm} \theta )  ~ , \]
\[  v^i \phi^m
(\theta \gamma^{lm} \theta ) 
(\theta \gamma^{ijk} \theta )
(\theta \gamma^{jkl} \theta )  ~ , \]
\[  v^i \phi^m
(\theta \gamma^{kl} \theta ) 
(\theta \gamma^{ijm} \theta )
(\theta \gamma^{jkl} \theta )  ~ , \]
\[ ( v^i \phi^i ) \phi^j \phi^k
(\theta \gamma^{jl} \theta ) 
(\theta \gamma^{kmn} \theta )
(\theta \gamma^{lmn} \theta )   ~ , \]
\[ ( v^i \phi^i ) \phi^j \phi^k
(\theta \gamma^{lm} \theta ) 
(\theta \gamma^{jln} \theta )
(\theta \gamma^{kmn} \theta )   ~ , \]
\[ ( v^i \phi^i )
(\theta \gamma^{jk} \theta ) 
(\theta \gamma^{jlm} \theta )
(\theta \gamma^{klm} \theta ) = 0  ~ . \]
Again, these terms either vanish or reduce to the terms of $7^{v J J
J}$ upon using the Fierz identities Eqs.~(\ref{fier1}),
(\ref{fier2}) and (\ref{fier3}).

\section{Fierz Identities}

An efficient algorithm for generating Fierz identities has recently
been given in Ref.~\cite{taylor}.  By implementing that algorithm
using Mathematica, we obtain the following Fierz identities.
\begin{equation}
  (\epsilon \gamma^{a_1 a_2} \theta ) 
   ( \theta \gamma^{a_1 a_2} \theta ) = 0 
\end{equation}
\begin{equation}
 (\epsilon \gamma^{a_1 a_2 a_3} \theta ) 
   ( \theta \gamma^{a_1 a_2 a_3} \theta ) = 0 
\label{fier1}
\end{equation}
\begin{equation} 
(\epsilon \gamma^{a_1 a_2 i} \theta ) 
   ( \theta \gamma^{a_1 a_2} \theta ) = 
2  (\epsilon \gamma^{a_1} \theta ) 
   ( \theta \gamma^{a_1 i} \theta ) 
\end{equation}
\begin{equation}
  (\epsilon \gamma^{a_1 a_2} \theta ) 
   ( \theta \gamma^{a_1 a_2 i} \theta ) = 
-2  (\epsilon \gamma^{a_1} \theta ) 
   ( \theta \gamma^{a_1 i} \theta ) 
\end{equation}
\begin{equation}
  (\epsilon \gamma^{a_1 a_2 a_3 i} \theta ) 
   ( \theta \gamma^{a_1 a_2 a_3} \theta ) = 
-6  (\epsilon \gamma^{a_1} \theta ) 
   ( \theta \gamma^{a_1 i} \theta ) 
\end{equation}
\begin{equation}
 (\theta \gamma^{a_1 a_2} \theta )
 (\theta \gamma^{a_1 a_2 i } \theta ) = 0
\end{equation}
\begin{equation}
(\theta \gamma^{a_1 a_2 i} \theta )
(\theta \gamma^{a_1 a_2 j} \theta ) = 
  2 (\theta \gamma^{a_1 i} \theta )
    (\theta \gamma^{a_1 j} \theta )
  			\label{fier2}
\end{equation}
\begin{equation}
 (\theta \gamma^{a_1 a_2 a_3} \theta )
 (\theta \gamma^{a_1 a_2 a_3 i j k } \theta ) = 0
\end{equation}
\begin{equation}
 (\theta \gamma^{a_1 i} \theta )
 (\theta \gamma^{a_1 j k } \theta )
+ (\theta \gamma^{a_1 j} \theta )
 (\theta \gamma^{a_1 k i } \theta )
+ (\theta \gamma^{a_1 k} \theta )
 (\theta \gamma^{a_1 i j } \theta ) = 0
\label{fier4}
\end{equation}
\begin{eqnarray}
(\theta \gamma^{a_1ij} \theta )
(\theta \gamma^{a_1kl} \theta ) &=&
  -3 (\theta \gamma^{ij} \theta )
     (\theta \gamma^{kl} \theta )
  -2 (\theta \gamma^{ik} \theta )
     (\theta \gamma^{jl} \theta )
  +2 (\theta \gamma^{il} \theta )
     (\theta \gamma^{jk} \theta )
			\nonumber \\
  & &
  - \delta^{jk} (\theta \gamma^{a_1 i} \theta )
                (\theta \gamma^{a_1 l} \theta )
  - \delta^{il} (\theta \gamma^{a_1 j} \theta )
                (\theta \gamma^{a_1 k} \theta )
  + \delta^{ik} (\theta \gamma^{a_1 j} \theta )
                (\theta \gamma^{a_1 l} \theta ) 
			\nonumber \\
  & & + \delta^{jl} (\theta \gamma^{a_1 i} \theta )
                    (\theta \gamma^{a_1 k} \theta )  
			\label{fier3}
\end{eqnarray}
\begin{eqnarray}
 0  &=&
    (\theta \gamma^{lm} \theta )
     (\theta \gamma^{ijk} \theta )
 -  (\theta \gamma^{km} \theta )
     (\theta \gamma^{ijl} \theta )
 +  (\theta \gamma^{kl} \theta )
     (\theta \gamma^{ijm} \theta )
			\nonumber \\
  & &
  - (\theta \gamma^{jm} \theta )
     (\theta \gamma^{ikl} \theta )
  + (\theta \gamma^{jl} \theta )
     (\theta \gamma^{ikm} \theta )
  - (\theta \gamma^{jk} \theta )
     (\theta \gamma^{ilm} \theta )
			\nonumber \\
  & &
  +  (\theta \gamma^{im} \theta )
     (\theta \gamma^{jkl} \theta )
  -  (\theta \gamma^{il} \theta )
     (\theta \gamma^{jkm} \theta )
  + (\theta \gamma^{ik} \theta )
     (\theta \gamma^{jlm} \theta )
  + 3 (\theta \gamma^{ij} \theta )
     (\theta \gamma^{klm} \theta )
			\nonumber \\
  & &  - \delta^{ik} (\theta \gamma^{a_1 j} \theta )
     (\theta \gamma^{a_1 lm} \theta )
 + \delta^{il} (\theta \gamma^{a_1 j} \theta )
     (\theta \gamma^{a_1 km} \theta )
 -  \delta^{im} (\theta \gamma^{a_1 j} \theta )
     (\theta \gamma^{a_1 kl} \theta ) \nonumber \\
   & &   + \delta^{j k} (\theta \gamma^{a_1 i} \theta )
     (\theta \gamma^{a_1 lm} \theta )
 - \delta^{jl} (\theta \gamma^{a_1 i} \theta )
     (\theta \gamma^{a_1 km} \theta )
 +  \delta^{jm} (\theta \gamma^{a_1 i} \theta )
     (\theta \gamma^{a_1 kl} \theta )
\label{fierfor6}
\end{eqnarray}
\begin{eqnarray}
(\theta \gamma^{a_1 a_2 i} \theta )
(\theta \gamma^{a_1 a_2 jklm} \theta )  &=&
   2 (\theta \gamma^{lm} \theta )
     (\theta \gamma^{ijk} \theta )
 - 2 (\theta \gamma^{km} \theta )
     (\theta \gamma^{ijl} \theta )
 + 2 (\theta \gamma^{kl} \theta )
     (\theta \gamma^{ijm} \theta )
			\nonumber \\
  & &
  + 2 (\theta \gamma^{jm} \theta )
     (\theta \gamma^{ikl} \theta )
 - 2 (\theta \gamma^{jl} \theta )
     (\theta \gamma^{ikm} \theta )
 + 2 (\theta \gamma^{jk} \theta )
     (\theta \gamma^{ilm} \theta )
			\nonumber \\
  & &  + 2 (\theta \gamma^{im} \theta )
     (\theta \gamma^{jkl} \theta )
 - 2 (\theta \gamma^{il} \theta )
     (\theta \gamma^{jkm} \theta )
 + 2 (\theta \gamma^{ik} \theta )
     (\theta \gamma^{jlm} \theta ) \nonumber \\
   & &  - 2 (\theta \gamma^{ij} \theta )
     (\theta \gamma^{klm} \theta )
\label{fier5}
\end{eqnarray}
\begin{eqnarray}
 0 &=&
 - 2 (\theta \gamma^{lm} \theta )
     (\theta \gamma^{ijk} \theta )
 + 2 (\theta \gamma^{km} \theta )
     (\theta \gamma^{ijl} \theta )
 - 2 (\theta \gamma^{kl} \theta )
     (\theta \gamma^{ijm} \theta )
			\nonumber \\
  & &
  - 2 (\theta \gamma^{jm} \theta )
     (\theta \gamma^{ikl} \theta )
  + 2 (\theta \gamma^{jl} \theta )
     (\theta \gamma^{ikm} \theta )
  - 2 (\theta \gamma^{jk} \theta )
     (\theta \gamma^{ilm} \theta )
			\nonumber \\
  & &  + 2 (\theta \gamma^{im} \theta )
     (\theta \gamma^{jkl} \theta )
 - 2 (\theta \gamma^{il} \theta )
     (\theta \gamma^{jkm} \theta )
 + 2 (\theta \gamma^{ik} \theta )
     (\theta \gamma^{jlm} \theta ) \nonumber \\
  & &  + 10 (\theta \gamma^{ij} \theta )
     (\theta \gamma^{klm} \theta )
 - (\theta \gamma^{a_1 a_2 j} \theta )
     (\theta \gamma^{a_1 a_2 iklm } \theta )
 +  (\theta \gamma^{a_1 a_2 i } \theta )
     (\theta \gamma^{a_1 a_2 jklm } \theta ) \nonumber \\
  & &  + 2 \delta^{ik} (\theta \gamma^{a_1 m} \theta )
     (\theta \gamma^{a_1 jl} \theta )
 - 2 \delta^{ik} (\theta \gamma^{a_1 l} \theta )
     (\theta \gamma^{a_1 jm} \theta ) 
 - 2 \delta^{il} (\theta \gamma^{a_1 m} \theta )
     (\theta \gamma^{a_1 jk} \theta ) \nonumber \\
  & &  +2  \delta^{il} (\theta \gamma^{a_1 k} \theta )
     (\theta \gamma^{a_1 jm} \theta )
 +2  \delta^{im} (\theta \gamma^{a_1 l} \theta )
     (\theta \gamma^{a_1 jk} \theta ) 
 - 2 \delta^{im} (\theta \gamma^{a_1 k} \theta )
     (\theta \gamma^{a_1 jl} \theta ) \nonumber \\
  & &  -2 \delta^{jk} (\theta \gamma^{a_1 m} \theta )
     (\theta \gamma^{a_1 il} \theta )
 +2  \delta^{jk} (\theta \gamma^{a_1 l} \theta )
     (\theta \gamma^{a_1 im} \theta ) 
 +2  \delta^{jl} (\theta \gamma^{a_1 m} \theta )
     (\theta \gamma^{a_1 ik} \theta ) \nonumber \\
  & &  -2 \delta^{jl} (\theta \gamma^{a_1 k} \theta )
     (\theta \gamma^{a_1 im} \theta )
 -2  \delta^{jm} (\theta \gamma^{a_1 l} \theta )
     (\theta \gamma^{a_1 ik} \theta ) 
 +2  \delta^{jm} (\theta \gamma^{a_1 k} \theta )
     (\theta \gamma^{a_1 il} \theta ) 
\label{fier6}
\end{eqnarray}

\end{document}